%% file: MVDR.tex
\documentclass[iop, twocolumn, twocolappendix, numberedappendix, appendixfloats]{emulateapj}

\pdfoutput=1
\usepackage[breaklinks,colorlinks=true,linkcolor=blue,anchorcolor=blue,citecolor=blue,urlcolor=blue,draft=false]{hyperref}
\usepackage{color, graphicx, amssymb, mathrsfs, placeins, times, xspace}

\usepackage{natbib}
\usepackage[utf8]{inputenc}

\newcommand{\percent}{\ensuremath{\%}}
\newcommand{\Msun}{M_\odot}
\newcommand{\Mobs}{M_\mathrm{obs}}
\newcommand{\Mgas}{M_\mathrm{gas}}
\newcommand{\Mbar}{M_\mathrm{bar}}
\newcommand{\Mstar}{M_\mathrm{star}}
\newcommand{\gbar}{\mathrm{g}_\mathrm{bar}}
\newcommand{\gobs}{\mathrm{g}_\mathrm{obs}}
\newcommand{\gdag}{\mathrm{g}_\mathrm{\dagger}}
\newcommand{\gddag}{\mathrm{g}_\mathrm{\ddagger}}
\newcommand{\fstar}{f_\mathrm{star}}
\newcommand{\fbar}{f_\mathrm{bar}}
\newcommand{\hi}{{\rm H\,{\footnotesize\rm I}}}
\newcommand{\Jinf}{J_{\infty}}

\shorttitle{MVDR in Galaxy Clusters}
\shortauthors{Tian et al.}

\begin{document}

%
%
\title{Mass-Velocity Dispersion Relation in HIFLUGCS Galaxy Clusters}
\author{Yong Tian\altaffilmark{1}}
\author{Po-Chieh Yu\altaffilmark{2}}
\author{Pengfei Li\altaffilmark{3}}
\author{Stacy S. McGaugh\altaffilmark{3}}
\author{Chung-Ming Ko\altaffilmark{1,4}}

\email{Corresponding author: stacy.mcgaugh@case.edu}
\email{cmko@astro.ncu.edu.tw}
\altaffiltext{1}{Institute of Astronomy, National Central University, Taoyuan 32001, Taiwan}
\altaffiltext{2}{College of General Studies, Yuan-Ze University, Taoyuan 32003, Taiwan}
\altaffiltext{3}{Department of Astronomy, Case Western Reserve University, Cleveland, OH 44106, USA}
\altaffiltext{4}{Department of Physics and Center for Complex Systems, National Central University, Taoyuan 32001, Taiwan}

%
%
\begin{abstract}
We investigate the mass-velocity dispersion relation (MVDR) in 29 galaxy clusters
 in the HIghest X-ray FLUx Galaxy Cluster Sample (HIFLUGCS).
We measure the spatially resolved, line-of-sight velocity dispersion profiles
 of these clusters,
 which we find to be mostly flat at large radii,
 reminiscent of the rotation curves of galaxies.
We discover a tight empirical relation between the baryonic mass $M_\mathrm{bar}$
 and the flat velocity dispersion $\sigma$ of the member galaxies, i.e. MVDR,
 $\log(M_\mathrm{bar}/M_\odot)=4.1^{+0.4}_{-0.4}\,\log(\sigma/\mathrm{km}\,\mathrm{s}^{-1})+1.6^{+1.0}_{-1.3}$,
 with the lognormal intrinsic scatter of $12^{+3}_{-3}\%$.
 The residuals of the MVDR are uncorrelated with other cluster properties
  like temperature, cluster radius, baryonic mass surface density, and redshift.
 These characteristics are reminiscent of the MVDR for individual galaxies,
  albeit about ten times larger characteristic acceleration scale.
 The cluster baryon fraction falls short of the cosmic value, exposing a problem:
  the discrepancy increases systematically for clusters of lower mass and lower baryonic acceleration.
\end{abstract}



%
%

\section{Introduction}

Kinematic scaling relations provide key tests and hints in understanding the dark matter problem.
For example, \cite{TF77} found a relation between galactic luminosity and \hi\ line-widths.
Later on, the baryonic Tully-Fisher relation (BTFR), $\Mbar\propto v^{4}$,
 was revealed as a tight relation in spiral galaxies \citep{McGaugh00, Verheijen01, McGaugh11, Lelli16, Lelli19};
Similarly, the mass-velocity dispersion relation (MVDR), $\Mbar\propto \sigma^{4}$,
 was expected in pressure supported systems such as galaxy clusters and elliptical galaxies,
 or known as the Baryonic Faber-Jackson Relation
 \citep[BFJR,][]{FJ76, Sanders10, Catinella12, Cappellari13, Aquino-Ortiz18, Barat19}.
These tight relations raise the critical issues of
 how the stochastic processes of galaxy formation can produce regularity
 in $\Lambda$ cold dark matter ($\Lambda$CDM) model by adapting the abundance matching relation
 \citep{Dutton10, DW15, DW17, WT18, Katz19}.
On the other hand, MOdified Newtonian Dynamics (MOND)
 directly implied these relations in its framework \citep{Milgrom83, SM02, FM12}.

Analogous to Kepler’s and Newton's laws, the kinematic scaling relations,
 e.g., BTFR and BFJR, are the counterpart of the effective dynamical relations.
For instance, in galactic systems, both relations can be implied by
  the low acceleration limit of the radial acceleration relation \citep[RAR,][]{McGaugh16}.
The RAR is a tight relation between two independent measurements:
 the observed radial acceleration, $\gobs=|\partial\Phi_{\mathrm{tot}}/\partial\,r|=v^{2}/r$, and the baryonic acceleration, $\gbar=G\Mbar(<r)/r^{2}$, where $\Phi_{\mathrm{tot}}$ is the total gravitational potential and $\Mbar(<r)$ is the enclosed baryonic mass within the radius $r$.
The low acceleration limit of the RAR gives $\gobs\simeq\,\sqrt{\gbar\gdag}$
 with the acceleration scale $\gdag=(1.20\pm0.02)\times 10^{-10}$\,m\,s$^{-2}$ \citep{McGaugh16, Lelli17, Li18, McGaugh18}.
It incorporates $v_{\infty}^4=G\Mbar\gdag$ (BTFR)
 as well as $\sigma^4\propto\,G\Mbar\gdag$ (BFJR).
In addition, the RAR is also confirmed in elliptical galaxies
 \citep{Lelli17, TK17, Rong18, Chae19, TK19}.
As one interpretation of the RAR, MOND implied the BTFR and the BFJR
 almost four decades ago \citep{Milgrom83}.

Given the tight scaling relations in galactic systems, it would be interesting to investigate them
 on the largest gravitationally bound systems, clusters of galaxies.
The baryonic mass of a galaxy cluster includes hot intracluster medium (ICM) and member galaxies.
Due to strong gravitational potential of the cluster,
 ICM is mainly composed of ionized gases,
 which emits X-ray radiation mostly through the bremsstrahlung process \citep{Lea73}.
However, the luminous mass of a galaxy cluster is not enough for accounting its gravity,
 as revealed by dynamical studies \citep{Zwicky33, Bahcall77, KB12, Overzier16, Rines16}
 and gravitational lensing \citep{KN11, Umetsu20}.
The discrepancy is considerably large so that MOND, though predicts stronger dynamical effects
 at low accelerations, still presents a residual missing mass in galaxy clusters
 \citep{Sanders99, Sanders03, PS05, FM12, Tian20}.
This indicates that the RAR must be different in the cluster scale.

More recently, \cite{Tian20} studied 20 galaxy clusters from Cluster Lensing And Supernova survey with Hubble \citep[CLASH,][]{Postman12}, and found a tight correlation between $\gobs$ and $\gbar$,
\begin{equation}
 \label{eq:ClusterRAR}
 \gobs\simeq\sqrt{\gbar\gddag}.
\end{equation}
 with a new acceleration scale $\gddag=(2.0\pm0.1)\times10^{-9}$ ms$^{-2}$.
In the CLASH RAR, $\gobs$ is probed by strong-lensing, weak-lensing shear-and-magnification data
 \citep{Umetsu16},
 while $\gbar$ is estimated by the distribution of X-ray gas mass \citep{Donahue14}
  and stellar mass \citep{Chiu18}.
This tight relation is in contradiction to \cite{CD20},
 which claimed no such relation in the cluster scale.
If the validity of the CLASH RAR is extended from gravitational lensing to dynamics,
 it will imply a MVDR in the galaxy cluster, i.e.
 $\sigma^4\propto G\Mbar\gddag$, with a new acceleration scale $\gddag$
 \citep[see section 4.3 in][]{Tian20}.

In galaxy clusters, the attempts on the MVDR were indirectly studied by $\Mgas - T$:
 the X-ray gas mass $\Mgas$ and the temperature $T$ \citep{Sanders94, Ettori04, Angus08, FM12}.
\cite{Sanders94} first found a rough correlation of $\Mgas\propto\,T^{2}$ from 20 rich galaxy clusters.
\cite{FM12} obtained a similar result by combining rich clusters \citep{RB02, Sanders03}
 and galaxy groups \citep{Angus08}.
Their results demonstrated a slope consistent with $\Mgas\propto T^{2}$ by
 the prediction of MOND, but not with the normalization.
By contrast, this slope is deviated from the conventional prediction $\Mgas\propto T^{2/3}$
 in $\Lambda$CDM paradigm \citep{FM12, McGaugh15}.

In the literature, the MVDR has never been clearly quantified in galaxy clusters.
Some studies focus on the X-ray luminosity-velocity dispersion relation
 \citep[e.g.][]{MZ98, XW00, MG01, Popesso05, Zhang11}
 as both quantities are directly measured.
To convert $M - T$ into $M - \sigma$
  requires an additional scaling relation $T - \sigma$,
with both $\Mgas$ and $T$ depending on X-ray observations.
The normalization of the intercept implies a different acceleration scale
 in galaxy clusters \citep{FM12}.
However, this new scale has never been estimated numerically.
On the other hand, \cite{Milgrom19a} found the acceleration scale
 in intermediate-richness galaxy groups is consistent with MOND.

In this work, we investigate the MVDR by directly studying the baryonic mass of clusters
 and the velocity dispersion of member galaxies in 29 HIFLUGCS clusters.
The paper was organized as follows.
In section \ref{Sec2:Data}, we describe the properties of the sample and the methods to the
 velocity dispersion.
In section \ref{Sec3:Res}, we analyze the MVDR by Bayesian statistics and present the residuals versus
 other cluster properties.
In section \ref{Sec4:Dis}, we compare our result with the CLASH RAR,
 test it on the dark matter model, and estimate the baryon fraction against
 the baryonic mass and the baryonic acceleration.
Throughout this paper, we assume a flat $\Lambda$CDM cosmology with $\Omega_\mathrm{m}=0.3$,
 $\Omega_\Lambda=0.7$, and a Hubble constant of $H_0 = 70$\,km\,s$^{-1}$\,Mpc$^{-1}$,
 in order to be consistent with the measurement in the HIFLUGCS.

%
%

\section{Data \& Methods}\label{Sec2:Data}

Studying the MVDR of galaxy clusters needs both the baryonic mass and the velocity dispersion:
 (1) the majority of the baryonic mass is dominated by the X-ray gas mass
 due to the strong gravitational potential of the clusters;
 (2) in hydrostatic equilibrium, the velocity dispersion of member galaxies can represent
 the kinematics of a whole cluster.

The analysis of velocity dispersion requires the relative line-of-sight (los)
 velocity of member galaxies to the cluster center.
In the CLASH RAR, the cluster center is found to be the brightest cluster galaxy (BCG)
 in CLASH sample \citep{Umetsu14, Umetsu16}.
BCGs are usually positioned at the geometric and kinematic center of
 the cluster, the central peak of the X-ray emission \citep{JF84, LM04}, and the minimum of gravitational potential well \citep{Zitrin12}.
Because of the MVDR implied by the CLASH RAR, the consistent result needs to adopt the BCG as
 the cluster center.

For our study, an appropriate candidate of cluster database should provide a wide variety
 of the X-ray gas mass, optical measurements of BCGs and member galaxies, and the analysis of an offset between BCG position and the X-ray flux-weighted cluster center.
To satisfy these requirements, we found the HIFLUGCS \citep{RB02}
 which contained 64 galaxy clusters selected from the \emph{ROSAT} All-Sky Survey \citep{Ebeling98, Bohringer00, Bohringer04}.
The X-ray measurement of 63 galaxy clusters was refined by combing the excellent quality X-ray data
 in the \emph{XMM-Newton} archive \citep{Zhang11}.
Besides, they collected member galaxies and analyzed the offset of clusters center.

In this work, we focus on a subset of 29 clusters in the HIFLUGCS by the constraints
 of the cluster center offset and the information of member galaxies.
The median of the offset in 61 HIFLUGCS clusters is 12 kpc ranging from 0.4 to 955 kpc.
Since a typical BCG effective radius $\approx 30$ kpc\citep{Schneider83, Schombert86, Tian20},
 we restrict the cluster center offset limited within 60 kpc, which narrow down to 51 galaxy clusters.
It is relatively small compared with the average HIFLUGCS cluster radius, $r_{500}\approx 1$ Mpc,
 at which the mass density is 500 times than the critical density.
Among these, only 29 clusters have the required information of member galaxies in the literature.
The properties of 29 BCGs and clusters are listed in Table~\ref{tab:BCGs} and ~\ref{tab:clusters}, respectively.

\input{Table1.tex}

\input{Table2.tex}

    \subsection{Baryonic Mass}
The baryonic mass $\Mbar$ of galaxy clusters comprises X-ray gas mass $\Mgas$ as the major component
 and the stellar mass $\Mstar$ as a minor one.
Most baryons in clusters are in the form of the ionized gas emitting X-ray
 due to the strong gravitational potential of clusters.
The sub-dominated component includes the stellar mass of BCGs and all member galaxies.

\cite{Zhang11} found 63 clusters in the HIFLUGCS available in the \emph{XMM-Newton} archive.
They analyzed $\sim 1.3$ Ms \emph{XMM-Newton} data for 57 clusters excluding
 four flared clusters and two with multiple redshifts.
The X-ray gas mass was estimated within $r_{500}$ by combining both \emph{XMM-Newton}
 and \emph{ROSAT} X-ray data.
In our subset of HIFLUGCS clusters, the X-ray gas mass ranges from $1.7\times10^{11} \Msun$ to $1.3\times10^{14} \Msun$, which are listed in Table \ref{tab:clusters}.

Apart from the dominated component of the X-ray gas mass, the total baryonic mass still needs
 the small contribution of the stellar mass.
The fraction between the stellar mass and the baryonic mass, $\fstar(r)\equiv\Mstar(<r)/\Mgas(<r)$,
 depends on gas mass and radius.
For a large cluster with the typical X-ray gas mass $\sim5\times10^{14}\Msun$, it is
  negligible ($\sim6\percent$) at one Mpc \citep{Chiu18} which is the average HIFLUGCS cluster radius of $r_{500}$ in our subsample.
On the other hand, this fraction increases towards lower gas mass.
For example, \cite{KB12} found it around $5-20\percent$ at $r_{500}$ corresponding to $\Mgas$
 ranging in $10^{14}-10^{13}\Msun$.

We estimated the stellar mass by the scaling relation of the fraction $\fstar$ at $r_{500}$ \citep[e.g., see their Equation~(11) in][]{Giodini09}.
With 91 COSMOS X-ray selected clusters and 27 nearby X-ray clusters, \cite{Giodini09} found
    \begin{equation}\label{eq:stellar mass}
    \fstar=(5.0^{+0.1}_{-0.1})\times10^{-2}\left(\frac{\Mgas}{5\times10^{13}\Msun}\right)^{-0.37^{+0.04}_{-0.04}}\,.
    \end{equation}
According to this relation, the median of $\fstar$ is $7\percent$ for our sample.
However, it is significant for two smallest clusters: Fornax ($\fstar=27\percent$)
 and NGC 4636 group ($\fstar=41\percent$).
Therefore, we obtain the baryonic mass of 29 HIFLUGCS clusters as well as
 their uncertainties, as listed in Table~\ref{tab:clusters}.

    \subsection{Velocity Dispersion}
Implied by the CLASH RAR, the flat velocity dispersion of the cluster in the tail corresponds to
 the total baryonic mass for the MVDR.
Assuming hydrostatic equilibrium, member galaxies are the tracers of the gravitational potential
 in the galaxy clusters.
We can study the los velocity of the member galaxies to get
 a flat velocity dispersion of the cluster by treating the BCG as the cluster center.

    \subsubsection{BCGs and Member Galaxies}

    \begin{figure*}[!htb]
    \centering
    \includegraphics[width=2.0\columnwidth]{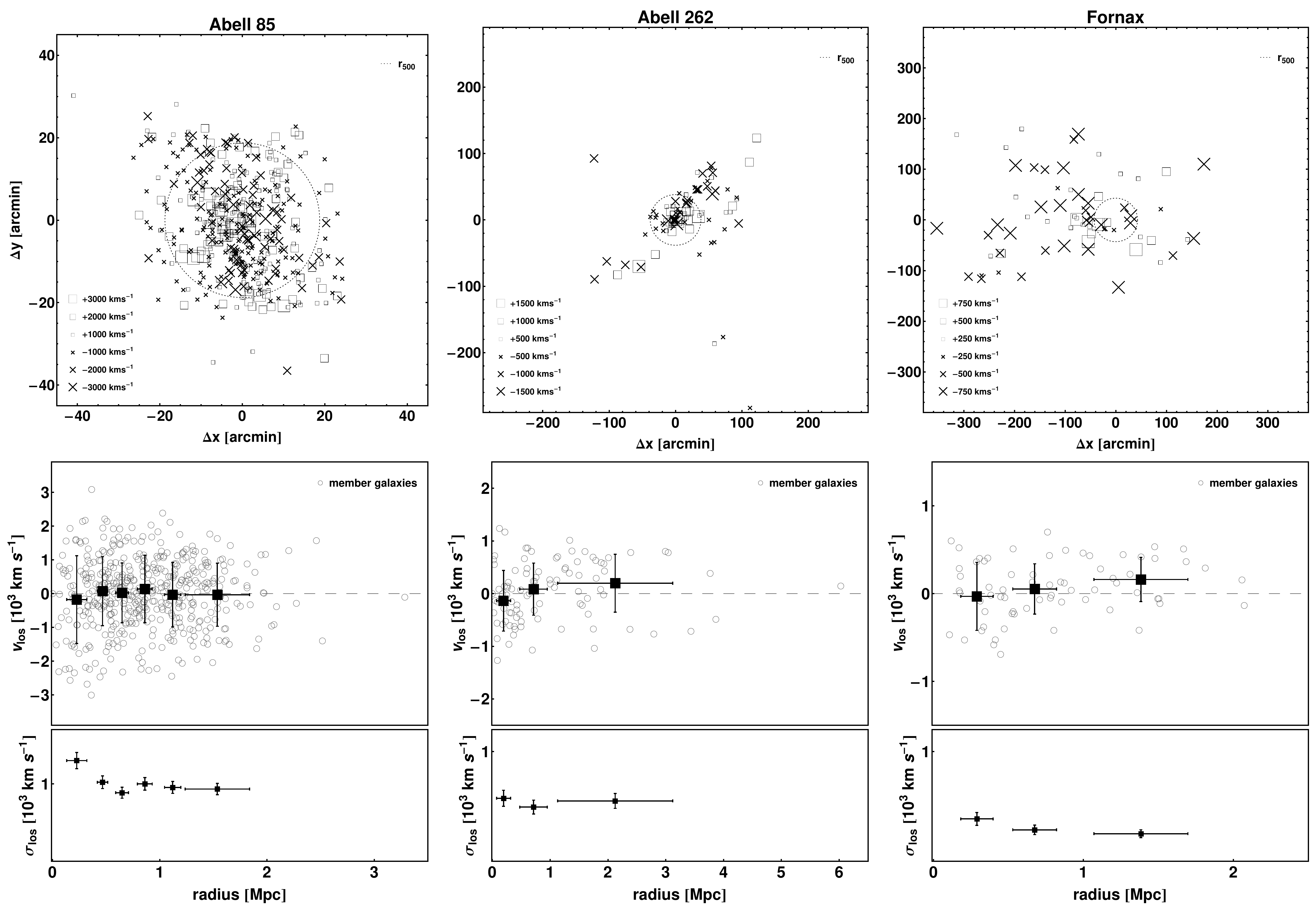}
    \caption{
Three examples (Abell 85, Abell 262, Fornax) of 29 HUFLUGCS clusters.
Abell 85 is the most massive among the three while Fornax is the least massive.
Upper panel: The spatial distributions and the los velocity of member galaxies are relative to the center point of BCG.
The symbol sizes are proportional to the velocity magnitudes.
Crosses indicate receding velocities; boxes, approaching velocities; dotted circles, $r_{500}$.
Middle panel: the relative los velocity ($V_{\mathrm{los}}$) distribution is in terms of the projected radius
 relative to the BCG.
The black filled rectangles represent the mean of the relative los velocity in each bin.
The los velocity dispersion ($\sigma_{\mathrm{los}}$) is the vertical standard deviation of the binned data.
Lower panel: the los velocity dispersion ($\sigma_{\mathrm{los}}$) present a flat tail for each cluster.
    }\label{fig:1}
    \end{figure*}

We collected BCGs and member galaxies from the literature,
 which was organized and identified by the probability of membership in SIMBAD\footnote{http://simbad.u-strasbg.fr/simbad/} \citep{SIMBAD}.
We carefully excluded uncertain members as well as the incomplete optical data of galaxies.
As the memberships identified by multiple references, we eliminate the repeated members.
Information of BCGs and member galaxies include the position, los velocity, and redshift.
Three examples (Abell 85, Abell 262, Fornax) of the two-dimensional distributions are present
 in the upper panel of Figure~\ref{fig:1}.

By requiring the member galaxies more than 30, we had the subsample of 29 clusters
 in the HIFLUGCS besides the Abell 2199 supercluster.
In total, there are 4926 member galaxies in our sample, which gives 170 members per cluster on average.
Among these, we excluded the Abell 2199 supercluster because of its complexity of multiple BCGs
 and clusters \citep{Lee15L}.
Finally, all available candidates are remained within the 29 galaxy clusters in our study.

Most clusters in the HIFLUGCS have a single BCG in their center, but some have a dumbbell,
 pair BCGs, or even multiple nuclei (A2199).
However, \cite{Zhang11} reported 13 HIFLUGCS clusters with more than one BCG.
In those cases, the BCG position is considered at the brighter one or the center between the dumbbell.
In our sample, five of them (A576, A2634, A3158, A3266, A3391) have dumbbell BCGs.
We chose either one of them as the cluster center.

In our sample, most of the member galaxies are identified by the optical measurement.
However, beside 234 members from the optical measurement in galaxy group NGC 4636
 (or referred as M31 group), 12 members are determined by using \hi\ measurement \citep{Kilborn09}.
In the velocity distribution diagram, these 12 members form a separated group
 from the other 234 members.
In addition, the relative los velocity of them is systematically deviated from the BCG.
Thus, we excluded these 12 galaxies from the membership of NGC 4636 group.

    \subsubsection{Methods}
The process of estimating the velocity dispersion of a galaxy cluster is similar
 to that of an elliptical galaxy, i.e. to treat a cluster as a galaxy and
 member galaxies as stars.
With the BCG as the center, we compute the relative projected radius and
 relative los velocity for each member galaxy,
 see three examples in the lower panel of Figure~\ref{fig:1}.
In our sample, the BCGs are sitting at the geometric and kinematic center,
 which justifies our requirements.

Because the velocity dispersion measured within larger radii
 is better to represent the total kinetic energy,
 we calculate them by the binned data of the member galaxies.
For each bin, the histogram presents a Gaussian distribution in the velocity profile.
The los velocity dispersion is evaluated by the half Gaussian width of
 the relative los velocity profile.
We also analyzed the error of the los velocity dispersion.
All 29 clusters show a flat velocity dispersion in the tail.
We adopted the last bin for the los velocity dispersion,
 which is listed in Table~\ref{tab:clusters}.

To evaluate non-Gaussian effects,
  we implement
  the biweight estimator \citep{Beers90}
  as an independent, statistically robust method to determine the velocity dispersion.
The biweight scale for the last binned data point
 yields highly consistent results.
Because the last binned data are presented as the Gaussian distribution,
 the biweight scale asymptotically approaches the Guassian method.

We compare our results with the total velocity dispersion reported by
 \cite{Zhang11} .
We found no distinguishable differences in the total velocity dispersions except for A3526,
 which is estimated as $890$ km/s instead of $486$ km/s \citep[their Table 1 in][]{Zhang11}.
Without their dataset of memeber galaxies,
 we can't check this discrepancy further.
Despite that, our calculation is more reasonable
 because the velocity dispersion is expected to be propotional to  X-ray gas mass.
As for the flat tail, half of them are almost identical to
 the total velcoity dispersion.
The rest of them is $\sim$10$\percent$ smaller than the total value.
This discovery is also similar to theirs.

%
%

\section{Results}\label{Sec3:Res}

Our main goal is to explore the empirical kinematic scaling relation
 between two independent measurements:
 the baryonic mass of galaxy clusters and the flat los velocity dispersion of the member galaxies,
 i.e., the MVDR.
To understand the tightness of the relation and the correlations among the cluster properties,
 we also study the intrinsic scatter and the residuals with Bayesian statistics.
The relation and the residuals are presented in the following subsections.

\subsection{MVDR with Bayesian Statistics}
In the logarithmic plane of the MVDR,
 29 HIFLUGCS clusters are distributed as a linear relation.
We model it by $y=mx+b$ with two independent variables: $y\equiv\ln(\Mbar/\Msun)$
 and $x\equiv\ln(\sigma_{\rm los}/\mathrm{km}\,\mathrm{s}^{-1})$.
With Bayesian statistics,
 we adopt the orthogonal-distance-regression (ODR) method
 with a Markov Chain Monte Carlo (MCMC) analysis in \cite{Lelli19}.
The ODR method is reasonable as two criteria are satisfied:
(1) a Gaussian intrinsic scatter perpendicular to the fitting line;
(2) two independent errors of x-axis $\sigma_{x_i}$ and y-axis $\sigma_{y_i}$.

    The log-likelihood function is written as
\begin{equation}\label{eq:log-likelihood}
     -2\ln\,\mathcal{L}=\sum_{i}\,\ln{(2\pi\sigma^2_{i})}+\sum_{i}\,\frac{\Delta_i^2}{\sigma^2_{i}}\,,
\end{equation}
    with
\begin{equation}\label{eq:Delta}
    \Delta_i^2=\frac{(y_i-m\,x_i-b)^2}{m^2+1}\,,
\end{equation}
    where $i$ runs over all data points, and $\sigma_i$ includes the
    observational uncertainties $(\sigma_{x_i}, \sigma_{y_i})$ and
    the lognormal intrinsic scatter $\sigma_\mathrm{int}$ \citep[e.g., see APPENDIX A in][]{Lelli19},

\begin{equation}\label{eq:sigma}
     \sigma^2_{i}=\frac{m^2\sigma^2_{x_i}}{m^2+1}+\frac{\sigma^2_{y_i}}{m^2+1}+\sigma_\mathrm{int}^2\,.
\end{equation}

\begin{figure*}[!htb]
    \centering
    \includegraphics[width=0.85\columnwidth]{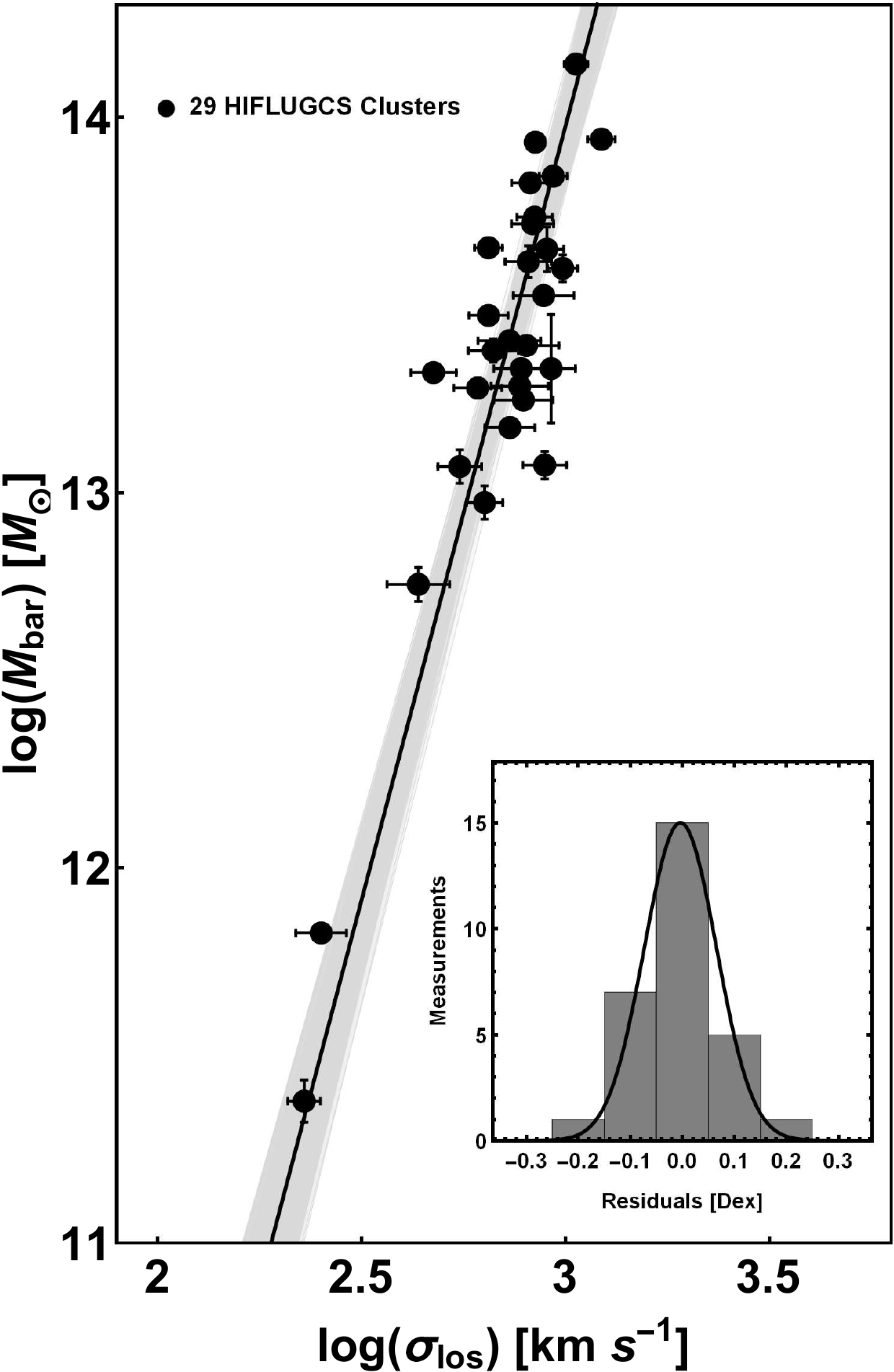}
    \includegraphics[width=1.20\columnwidth]{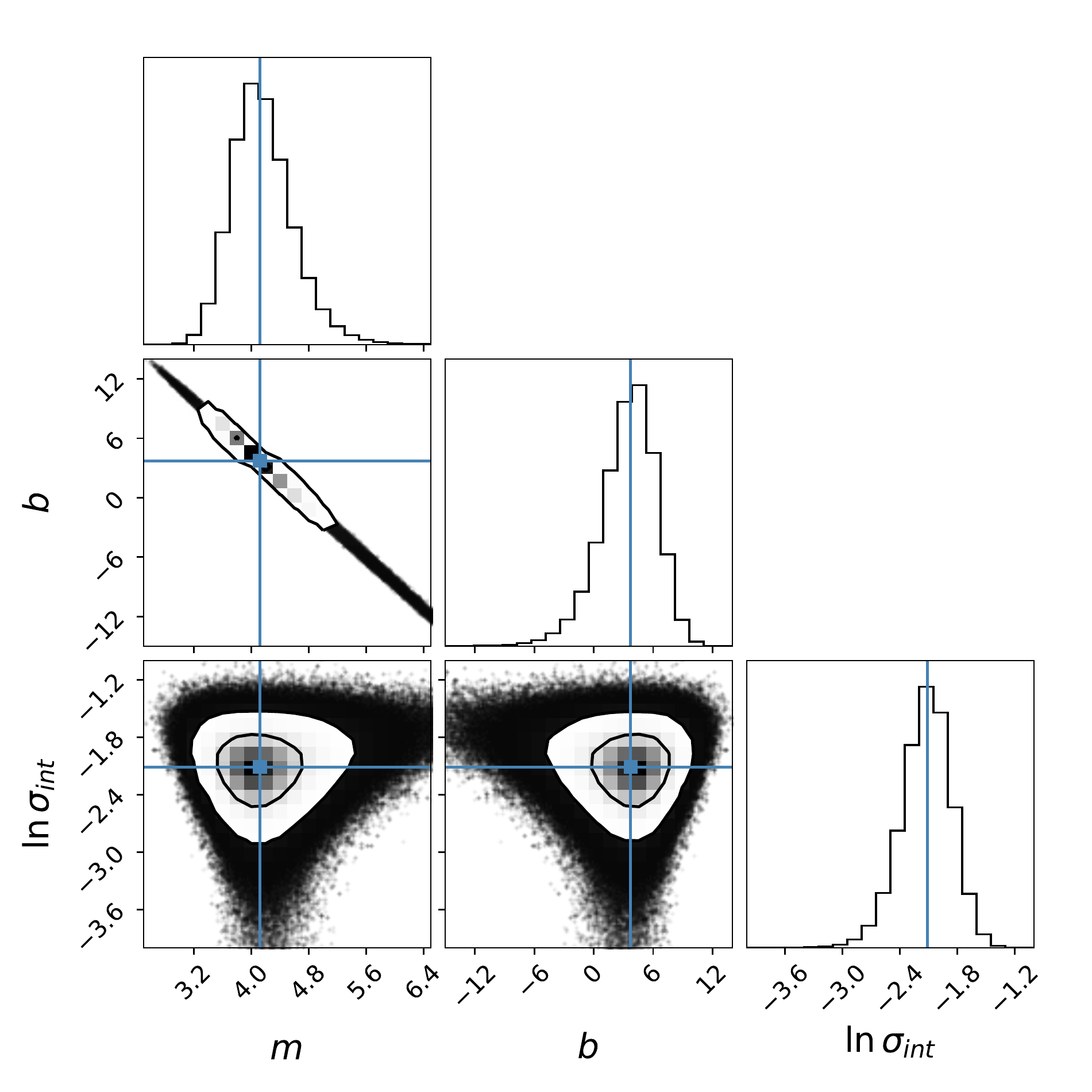}
    \caption{
The MVDR in 29 HIFLUGCS galaxy clusters.
Left panel: the kinematic scale relation between the baryonic mass $\Mbar$ and
 the flat los velocity dispersion $\sigma_{\rm los}$ of member galaxies.
The black line represent the best-fitting relation with MCMC,
 $\log(\Mbar/\Msun)=4.1^{+0.4}_{-0.4}\log(\sigma_{\rm los}/\mathrm{km}\,\mathrm{s}^{-1})+1.6^{+1.0}_{-1.3}$.
The gray shaded area represents one sigma region around the best-fitting black solid line.
The inset panel is a histogram of the orthogonal residuals presenting
 as a Gaussian distribution.
Right panel: constraints on the regression parameters for the MVDR
 with marginalized one-dimensional (histograms) and two-dimensional posterior distributions.
    }\label{fig:2}
\end{figure*}

We implement MCMC analysis for the slope and the intercept of the MVDR by EMCEE
 \citep{Foreman-Mackey13, emcee2019}.
We use non-informative flat priors on the slope $m$ and the intercept $b$
 within the interval of $[-100, 100]$, and the intrinsic scatter $\ln\,\sigma_{\rm int}\,\in [-5,2]$, see the result in Figure~\ref{fig:2}.
It gives a relation as
\begin{equation}\label{eq:MVDR}
    \log\left(\frac{\Mbar}{\Msun}\right)=4.1^{+0.4}_{-0.4}\log\left(\frac{\sigma_{\rm los}}{\mathrm{km}\,\mathrm{s}^{-1}}\right)+1.6^{+1.0}_{-1.3}\,,
\end{equation}
which is a tight relation with the error budget of the lognormal intrinsic scatter of
 $12^{+3}_{-3}\percent$.
Our result is consistent with the CLASH RAR, which implies $\Mbar\propto\sigma^{4}$.

To justify the initial assumption on a Gaussian intrinsic scatter perpendicular
 to the fitting line,
 we examine the histogram of the orthogonal residuals $\Delta_i$ with respect to Equation~\ref{eq:MVDR} (see the inset panel of Figure \ref{fig:2}).
The distributions of the residuals presents a Gaussian distribution
 with a tiny half width (0.07 dex).

To compare the new acceleration scale $\gddag$ with Equation~(\ref{eq:ClusterRAR}),
 the slope must be fixed to four.
This scale $\gddag$ depends on the intercept which is highly sensitive to the slope,
 see the posterior distributions of the fitting parameters in Figure~\ref{fig:2}.
In addition, the correct unit of the acceleration scale demands the exactly four of the slope.
We perform MCMC analysis again to get
\begin{equation}\label{eq:MVDR m=4}
    \log\left(\frac{\Mbar}{\Msun}\right)=4\log\left(\frac{\sigma_{\rm los}}{\mathrm{km}\,\mathrm{s}^{-1}}\right)+1.96^{+0.05}_{-0.06}\,,
\end{equation}
 which is still tight with the lognormal intrinsic scatter of $12^{+3}_{-3}\percent$.
The uncertainty of the intercept is dramatically reduced for fixed slope.
The reason is the degeneracy in the $m$--$b$ diagram is broken.

Without the correction with the stellar mass, we also study the scaling relation of
 $\Mgas$ and $\sigma_{\rm los}$ by the ODR MCMC method, which gives
\begin{equation}\label{eq:gasMVDR}
    \log\left(\frac{\Mgas}{\Msun}\right)=4.3^{+0.5}_{-0.4}\log\left(\frac{\sigma_{\rm los}}{\mathrm{km}\,\mathrm{s}^{-1}}\right)+1.0^{+1.1}_{-1.3}\,.
\end{equation}
Although $\fstar$ is relatively large for two smallest clusters, the resulting difference
 in the slope changes from 4.1 to 4.3.
Moreover, the relation is still tight with the lognormal intrinsic scatter of
 $12^{+3}_{-3}\percent$.

Two low-mass clusters dominate the slope and the intercept.
Excluding these two points, we get a different
MVDR with a larger scatter:
$m = 5.4^{+2.2}_{-1.1}$ and $b = -2.0^{+3.2}_{-6.3}$.
This is similar to the importance of low surface brightness disks to the BTFR,
 smaller galaxy clusters play a major role in the MVDR.

Although the flat velocity dispersion is more representive for the total kinetic energy,
 we still check the MVDR with velocity dispersion measured at $r_{500}$
 due to the baryonic masses estimated within $r_{500}$.
This makes a difference as the velocity dispersion profiles in half of the sample
 have not flattened yet at $r_{500}$.
It yields a slightly steeper MVDR: $m = 4.5^{+0.5}_{-0.4}$
 and $b = 0.6^{+1.1}_{-1.4}$.

    \subsection{Residuals}
\begin{figure*}[!htb]
    \centering
    \includegraphics[width=2.0\columnwidth]{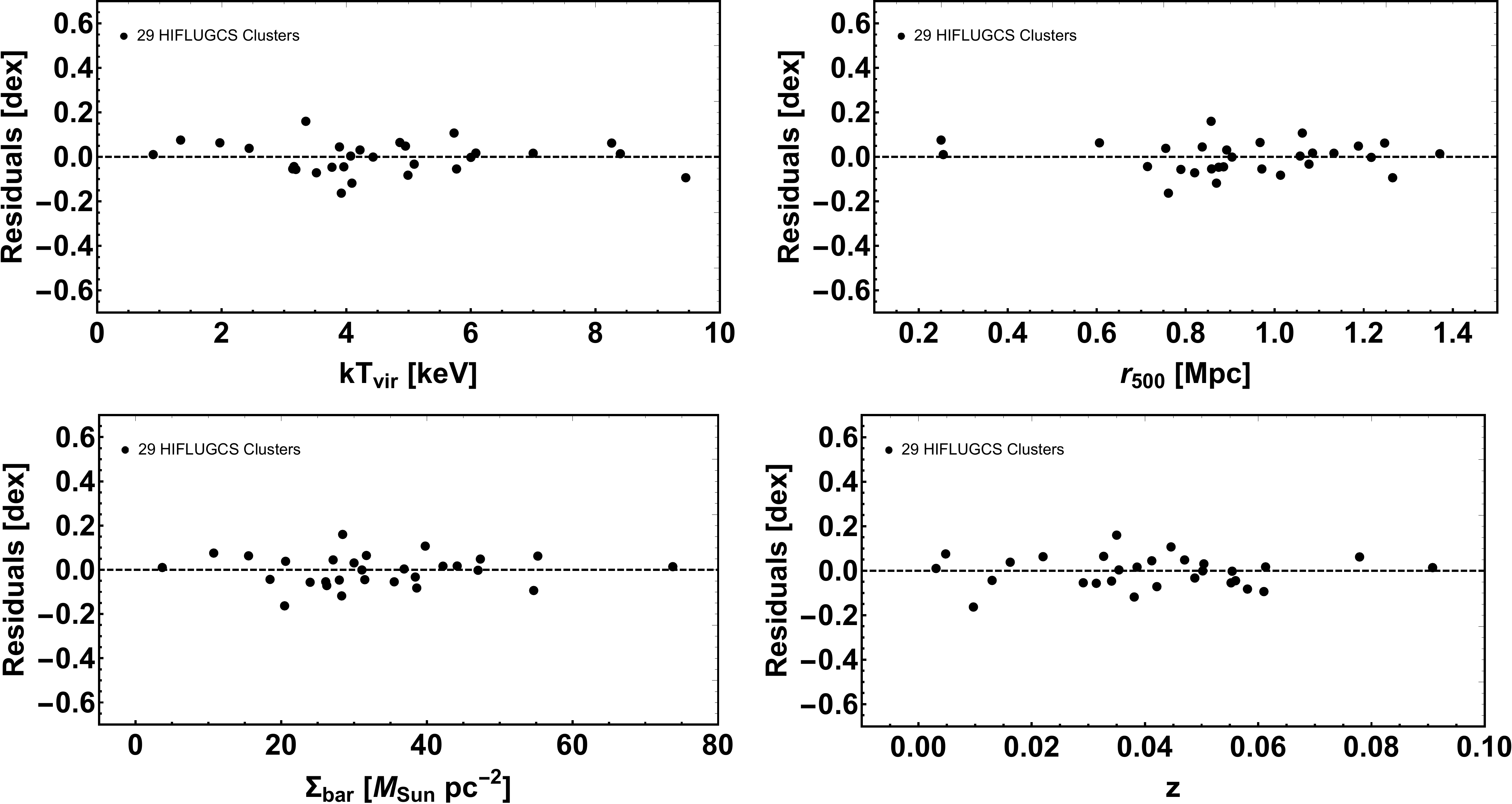}
    \caption{The residuals of the MVDR in 29 HIFLUGCS clusters.
The orthogonal residuals after subtracting Equation~(\ref{eq:MVDR}) against four cluster properties:
 virial temperature $k T_{\rm vir}$ (upper-left), cluster radius $r_{500}$ (upper-right), baryonic mass surface density $\Sigma_{\mathrm{bar}}$ (lower-left), and redshift (lower-right).
The dashed line represents zero difference.
    }\label{fig:3}
\end{figure*}

We present the orthogonal residuals by considering $\Delta_i$ in Equation~(\ref{eq:Delta})
 with $m$ and $b$ in Equation~(\ref{eq:MVDR})
 versus four cluster quantities: virial temperature $k T_{\rm vir}$, cluster radius $r_{500}$, baryonic mass surface density $\Sigma_{\mathrm{bar}}$,
 and redshift, see Figure~\ref{fig:3}.
The tiny residuals range between $-0.2$ to $0.2$ and display no significant correlation
 with four cluster properties.
These properties are reminiscent of the BTFR for individual galaxies.

%
%
\section{Discussions}\label{Sec4:Dis}


\subsection{Implications by the CLASH RAR}
\cite{Tian20} have revealed a tight RAR by 20 CLASH BCGs and clusters as
 $\gobs\simeq\sqrt{\gbar\gddag}$.
If we related $\gobs(r)$ to the velocity dispersion as $\gobs(r)=J(r)^{1/2}\sigma_{\mathrm{r,3D}}^2/r$
 and $\gbar(r)=G\Mbar(<r)/r^2$
 into the CLASH RAR, we can recover the MVDR: $\Mbar=\sigma_{\mathrm{r,3D}}^{4}J(r)G^{-1}\gddag^{-1}$.
Here, the Jeans factor $J(r)$ is related to the density profile and the
 anisotropic parameter $\beta$ by Jeans equation \citep[e.g., see][]{Milgrom84, Sanders10, FM12}:
\begin{equation}\label{eq:factorI}
    -[J(\tilde{r})]^{1/2}=\frac{d\ln\tilde{\rho}}{d\ln\,\tilde{r}}+2\beta\,,
\end{equation}
where $\tilde{r}\equiv\,r/r_{\ddagger}$ and $\tilde{\rho}\equiv\,\rho/\rho_{\ddagger}$ with
 the scale length $r_{\ddagger}=\sigma_{\mathrm{r,3D}}^2\gddag^{-1}$ and the scale density $\rho_{\ddagger}=\sigma_{\mathrm{r,3D}}^4(4\pi\,r_{\ddagger}^{3}G\gddag)^{-1}$.
Thus, as $\tilde{r}\rightarrow\infty$, we have $J(\tilde{r})\rightarrow\,\Jinf$ and
 $d\ln\tilde{\rho}/d\ln\,\tilde{r}\rightarrow\,-\alpha_{\infty}$ into $\Jinf=(\alpha_{\infty}-2\beta)^2$.

By the implications of the CLASH RAR, the MVDR is presented as
\begin{equation}\label{eq:MVDR_Jeans}
  \log(\Mbar)=4\log(\sigma_{\mathrm{r,3D}})+\log(\Jinf G^{-1}\gddag^{-1})\,.
\end{equation}
Here, the factor $\alpha_{\infty}$ depends on the density profile of the systems and $\beta=0$ for
 isotropic case.
Thus, the intercept involves $\Jinf$, the new acceleration scale $\gddag$,
 and the gravitational constant $G$.

For a special case of the isotropic parameter $\beta=0$
 and the flat velocity dispersion $\sigma_{\mathrm{r,3D}}=$ const.,
 the radial velocity dispersion in three dimensional $\sigma_{\mathrm{r,3D}}$
 in Equation~(\ref{eq:MVDR_Jeans}) is identical to
 the flat los velocity dispersion $\sigma_{\rm los}$.
To estimate $\sigma_{\rm los}$ from $\sigma_{\mathrm{r,3D}}$,
 we consider the surface density
 at the projected radius weighted $\sigma$ \citep[e.g., see chapter 4.2 in][]{BT08}.
Thus, it is trivial that $\sigma_{\rm los}=\sigma_{\mathrm{r,3D}}$ for this special case.

The two unknown factors $\gddag$ and $\Jinf$ in Equation~(\ref{eq:MVDR_Jeans})
 can be estimated by the intercept of Equation~(\ref{eq:MVDR m=4}).
For example, if we take $\alpha_{\infty}\in [3, 5]$ and $\beta=0$,
 then the corresponding Jeans factor $\Jinf\in [9, 25]$.
From the intercept, we estimate the new acceleration scale
 $\gddag=(0.8-2.2)\times10^{-9}$ m s$^{-2}$,
 which is about ten times larger than $\gdag$.
The actual value of $\gddag$ requires the precise measurement of
 $\langle \Jinf\rangle$ of the sample.
On the other hand, we can estimate the Jeans factor by adopting the new acceleration scale
 in the CLASH RAR, $\gddag=(2.0\pm0.1)\times10^{-9}$ m s$^{-2}$.
Comparing with Equation~(\ref{eq:MVDR m=4}) again, we get $\langle \Jinf\rangle=23.7\pm1.4$,
 and then $\langle\alpha_{\infty}-2\beta\rangle=4.86\pm0.14$.

In our calculation, we require the radius $r$ far away from
 the scale length $r_{\ddagger}$ to justify the asymptotic behavior of
 the Jeans factor $J(\tilde{r})\rightarrow\,\Jinf$.
With the new acceleration scale $\gddag=1.4\times10^{-9}$ m s$^{-2}$,
 we can estimate $r_{\ddagger}=\sigma_{\mathrm{last}}^2\gddag^{-1}$ in 29 HIFLUGCS clusters \citep[see the scale parameters in ][]{Milgrom84}.
Therefore, we get the average of $\langle r_{\ddagger}\rangle=14$ kpc,
 which is insignificant comparing with the average radius
 of last binned data in Table~\ref{tab:clusters},
 $\langle r_{\mathrm{last}}\rangle\approx1300$ kpc.
This result is in good agreement with our previous assumption $r\gg r_{\ddagger}$.

  \subsection{Test for the dark matter model}

A MVDR can be derived from $\Lambda$CDM under general considerations.
Following \citet{Mo98, Navarro00, McGaugh10, McGaugh12},
 we compare the MVDR and that expected in the $\Lambda$CDM model.
The enclosed total mass within this radius is given by
\begin{equation}
    M_{500} = \frac{4\pi}{3}(500\rho_{\rm crit})r_{500}^3,
    \label{eq:M500wr500}
\end{equation}
where $\rho_{\rm crit} = 3H_0^2/8\pi G$ is the critical density of the Universe.

To relate the DM mass with velocity dispersion,
 we consider the gravitational potential traced by the velocity dispersion in pressure supported systems as
\begin{equation}
    \frac{GM_{500}}{r_{500}^2}=\frac{J^{1/2}(r)\sigma^2_{r, 3D}}{r_{500}}\,.
    \label{eq:MVD_CDM}
\end{equation}
Together with the baryon fraction $\fbar = M_{\rm bar}/M_{500}$ at $r_{500}$,
 we derive
 the relation between the baryonic mass and the velocity dispersion in $\Lambda$CDM
\begin{equation}\label{eq:MVDR_CDM}
  \log(\Mbar)=3\log(\sigma_{\rm los})+\log\left(\frac{\sqrt{2}J^{3/4}\fbar}{\sqrt{500} G H_0}\right)\,.
\end{equation}
Here, with the flat velocity dispersion and the assumption of the isotropic parameter, we adopt $\sigma_{\rm los}=\sigma_{\mathrm{r,3D}}$.
Noticeably, the $\Lambda$CDM model predicts a slope of three,
 which does not agree with the observed value of four.

\begin{figure}[!htb]
    \centering
    \includegraphics[width=0.90\columnwidth]{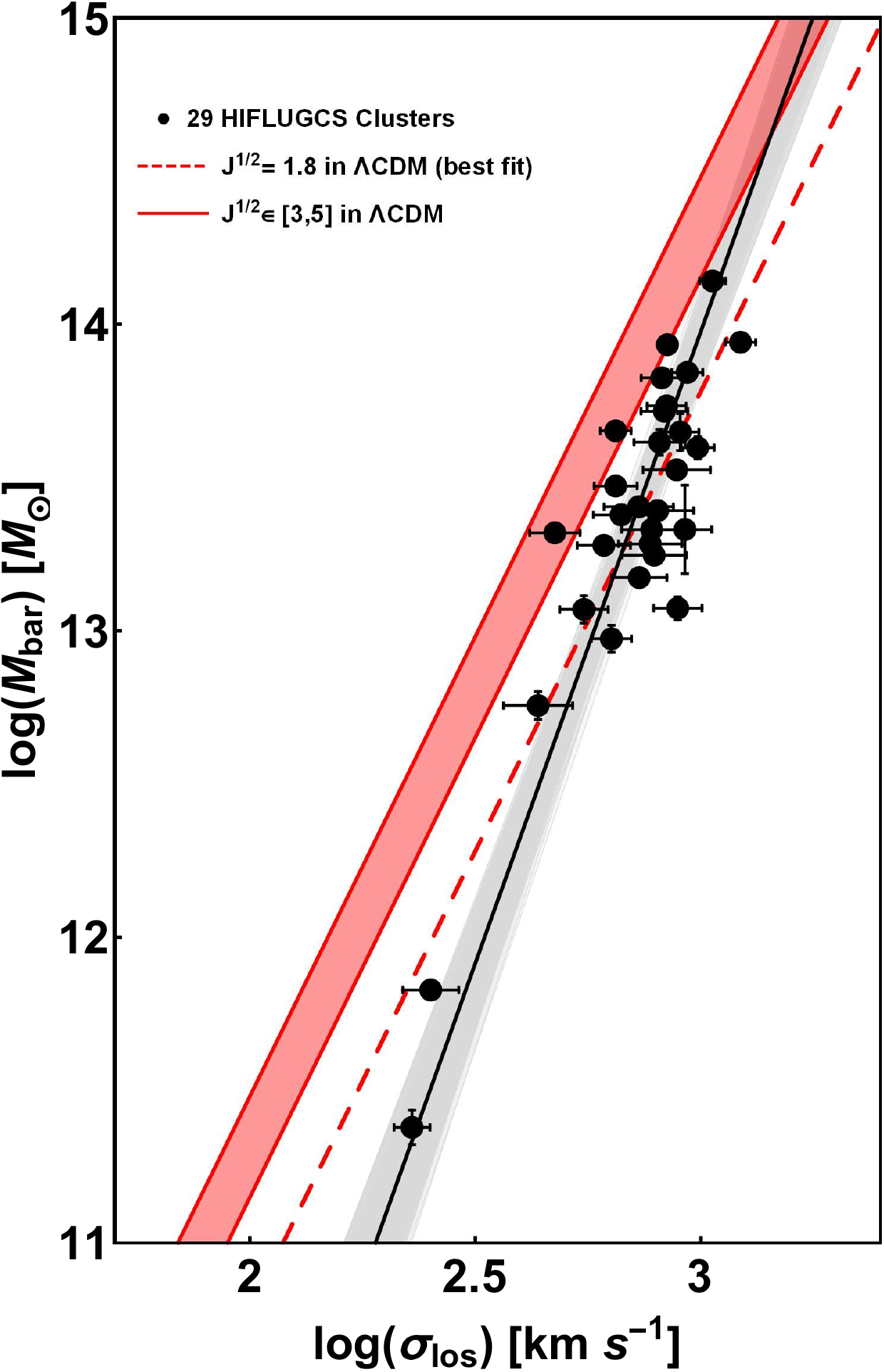}
    \caption{
Test for the dark matter model.
Black solid circles, black solid line, and gray shade area are the same as in Figure~\ref{fig:2}.
The red area and dashed line show the $\Lambda$CDM prediction
 with the choice of $J^{1/2}\in [3, 5]$
 and the best-fit of $J^{1/2}=1.8$, respectively.
     }\label{fig:4}
\end{figure}

The examination of the DM model requires the estimation of all the parameters
 of the intercept in Equation~(\ref{eq:MVD_CDM}):
 a Hubble constant, the baryon fraction, and the Jeans factor.
At $r_{500}$, we adopt $f_{\rm bar}=0.13$ measured by \cite{Donahue14, Tian20}.
While the Jeans factor depends on the density profile, we assume a possible range of $J^{1/2}\in[3,5]$.
A red region in Figure \ref{fig:4} shows the corresponding relation.
The $\Lambda$CDM prediction is systematically higher than the empirical MVDR,
 despite adopting a baryon fraction that is lower than the cosmic value.

To understand the best value of the Jeans factor,
 we implement the ODR MCMC technique for 29 HIFLUGCS clusters
 with the fixed slope ($m=3$), which gives
\begin{equation}\label{eq:DM_fit}
\log\left(\frac{\Mbar}{\Msun}\right)=3\log\left(\frac{\sigma_{\rm{los}}}{\mathrm{km}\,\mathrm{s}^{-1}}\right)+4.78^{+0.05}_{-0.06}\,.
\end{equation}
The intercept gives the best fit of $J^{1/2}=1.8$ shown as the red dashed line in Figure~\ref{fig:4}.
In this case, the velocity dispersion needs to be close to the circular velocity,
 which it is manifestly not.
As the best-fit Jeans factor is significantly lower than
plausible, it reveals a fine-tuning problem as discussed
in the next section.

In the literature, \cite{Mo98} predicted the dynamical MVDR
 in dark matter model as $M_{200}\propto V_{200}^3$.
Moreover, from both simulations and analytical relations, the results
 indicated $M_{200}\propto \sigma^{3}$ \citep{Evrard08, Rines13}.
To relate the dynamical mass $M_{200}$ with the baryonic mass $\Mbar$ requires
  specific model of the baryon faction.
If the baryon faction is a constant for all the cluster,
 dark matter prediction is inconsistent with the observed slope of four.

Observational results that obtain a slope close to three cover only a small dynamic range in mass.
Most studies of clusters only cover the range  $10^{14}<M_{200}<10^{15}\Msun$, and
 their results have not been tested for $M_{200}<10^{13}\Msun$. Since our sample
 covers a larger range in mass it provides a better constraint on the slope.
 Given the small number of groups with
  $M_{bar}<10^{12}\Msun$, further study will be important.

  \subsection{Implications for the baryon fraction}

\begin{figure}[!htb]
    \centering
    \includegraphics[width=1.0\columnwidth]{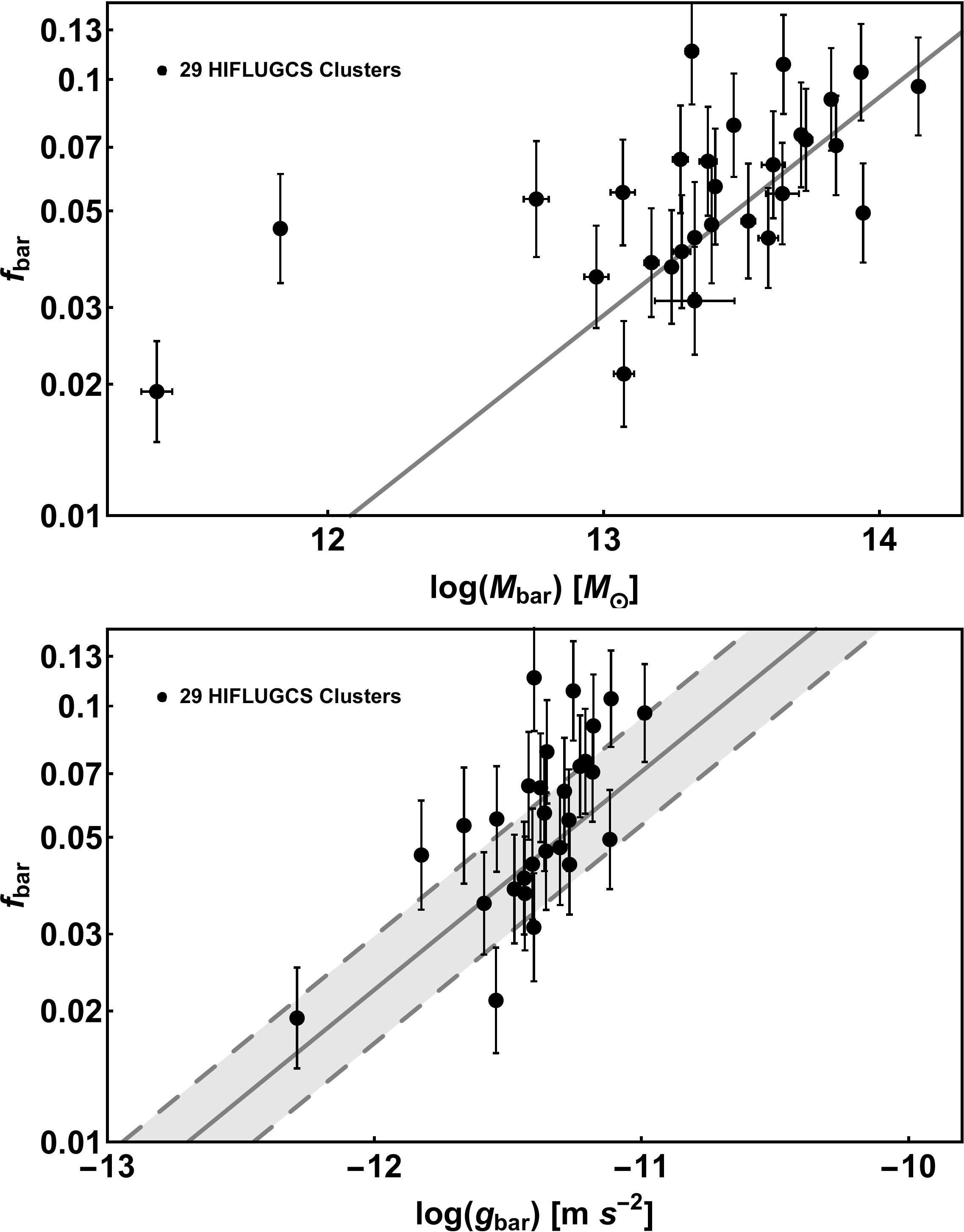}
    \caption{
The baryonic factions of the 29 HIFLUGCs Clusters (black solid circles) and that implied by the CLASH RAR (gray solid lines) are plotted against baryonic mass $M_{\rm bar}$ (upper panel) and baryonic acceleration g$_{\rm bar}$ (lower panel).
The dashed lines and the gray region are from the standard deviation of $r_{500}$.
    }\label{fig:5}
\end{figure}

In this Section, we study the baryon fraction $\fbar\equiv\Mbar/\Mobs=\gbar/\gobs$ of the 29 clusters.
We estimate $\fbar(r_{500})$ by considering the observational acceleration as
 $\gobs(r_{500})=\Jinf^{1/2}\sigma_{\mathrm{last}}^2/r_{500}$.
We consider $\sigma_{\rm last}\simeq\sigma_{500}$ given the flat velocity dispersion.
In addition, we adopt $\Jinf^{1/2}=4\pm1$.
The precise value of $\Jinf$ should vary among the sample inducing scatter.
The derived $f_{\rm bar}$ are plotted against baryonic mass (upper panel) and baryonic acceleration (lower panel) in Figure \ref{fig:5}.

For comparison, the CLASH RAR \citep{Tian20} implies
\begin{equation}
 \label{eq:fbar_gbar}
 \fbar(r)\approx\sqrt{\gbar(r)/g_\ddag}\,.
\end{equation}
Here, we estimate the baryonic acceleration at $r_{500}$
 by $\gbar(r_{500})=G\Mbar(<r_{500})/r_{500}^2$.
Since $r_{500}$ varies in every HIFLUGCS cluster, it contributes to the scatter of the relation.
We demonstrate Equation~(\ref{eq:fbar_gbar}) with the mean of
 $\langle r_{500}\rangle=917$ kpc and the error propagation to $\gbar(r_{500})$ from
 the standard deviation of $\sigma_{r_{500}}=259$ kpc,
 see gray solid and dashed line in Figure~\ref{fig:5}.
To compare with Equation~(\ref{eq:fbar_gbar}), we assume $\gddag=2.0\times10^{-9}$
 m s$^{-2}$ as found in \cite{Tian20}.

The 29 HIFLUGCS clusters are consistent with the implication of the CLASH RAR in Figure~\ref{fig:5}.
The relation of the baryon fraction with acceleration is reminiscent of MOND-like behavior ($\fbar\propto\gbar^{1/2}$), albeit with a different acceleration scale than in galaxies \citep{McGaugh04, McGaugh15, Milgrom19b}.

 \section{Summary}\label{Sec5:Sum}

In this paper, we investigate the correlation between
 the baryonic mass of 29 HIFLUGCS clusters and
 the flat velocity dispersion of their member galaxies.
We have calculated the baryonic mass by combining the X-ray gas mass \citep{Zhang11} and
 the stellar mass correction from \cite{Giodini09}.
By spatially resolving the velocity dispersion profiles,
 we find a flat velocity dispersion at large radii.
The resulting MVDR is well-described by a power-law relation,
 $\Mbar\propto\sigma_{\rm los}^{4.1\pm0.4}$,
 with a lognormal intrinsic scatter of $12^{+3}_{-3}\percent$.
This is consistent with the implication of the CLASH RAR, $\gobs\simeq\sqrt{\gbar\gddag}$.
Furthermore, the residuals do not correlate with other cluster properties
 such as temperature, cluster radius $r_{500}$, baryonic surface density, and redshift.

The MVDR can be naturally explained in the MOND paradigm.
The slope of four is a prediction of MOND.
However, the intercept of the MVDR indicates an acceleration scale $\gddag$ that is larger than that in galaxies.
Why clusters should exhibit MOND-like behavior with a shifted acceleration scale is a mystery in any theory.
In pure MOND, there should be no shift.
In terms of dark matter, there is no reason to expect MOND-like behavior at all.
One conceivable explanation is a dependence of the action on the depth potential well in addition to the acceleration \citep{ZF12}.

In this paper, we have derived the expected MVDR in $\Lambda$CDM from general considerations and
 examined the baryon fraction in our sample.
We find that $\Lambda$CDM gives a MVDR with a slope of three,
when we assume the Jeans factor and the baryon fraction are independent of the baryonic mass.
This is in serious tension with the observed slope of four.
Furthermore, the predicted intercept of the MVDR is systematically higher than observed.
This indicates a residual mass problem in $\Lambda$CDM:
 the predicted bayonic masses of galaxy clusters are more than required by the observed kinematics.
Possibly, this discrepancy could be compensated by
 introducing a systematic dependence of the Jeans factor and
 the baryon fraction on the baryonic mass of the clusters,
 but this creates a fine-tuning problem to get the slope right without
 inducing too much scatter.

We have revealed a tight MVDR on the cluster scale.
To explore the full range of this relation requires more data from smaller galaxy clusters.
This is also imperative for the understanding of the CLASH RAR in the small acceleration regime.

\section*{ACKNOWLEDGEMENTS}
We thank M. Milgrom for many inspiring discussions on this work,
and the anonymous referee for the valuable comments to improve the clarity of this paper.
YT and CMK are supported by the Taiwan Ministry of Science and Technology grant
 MOST 108-2112-M-008-006 and MOST 109-2112-M-008-005.
PCY is supported by the Taiwan Ministry of Science and Technology grant
 MOST 109-2112-M-155-001.
SSM and PL are supported in part by NASA ADAP grant 80NSSC19k0570 and NSF PHY-1911909.

\bibliographystyle{yahapj}
\bibliography{reference}

\end{document}

%% file: Table1.tex
\begin{deluxetable}{lcccr}
        \tablecolumns{5}
        \tabletypesize{\footnotesize }
        \tablewidth{\columnwidth}
        \tablecaption{\label{tab:BCGs}
        Properties of 29 HIFLUGCS BCGs
        }
        \tablehead{
         \multicolumn{1}{c}{Name} &
         \multicolumn{1}{c}{$z$} &
         \multicolumn{1}{c}{R.A.} &
         \multicolumn{1}{c}{decl.} &
         \multicolumn{1}{c}{$V_\mathrm{los}$\tablenotemark{a}} \\
         \colhead{}
         & &
         \multicolumn{1}{c}{(J2000.0)} &
         \multicolumn{1}{c}{(J2000.0)} &
         \multicolumn{1}{c}{(km/s)}
         }
         \startdata
NGC 4636	&	0.0031	&$	12	:	42	:	49.87	$&$	+02	:	41	:	16.01	$&	919	\\
Fornax	&	0.0048	&$	03	:	38	:	29.00	$&$	-35	:	27	:	02.67	$&	1422	\\
Abell 3526	&	0.0097	&$	12	:	48	:	49.28	$&$	-41	:	18	:	39.92	$&	2904	\\
Abell 1060	&	0.0130	&$	10	:	36	:	42.82	$&$	-27	:	31	:	42.02	$&	3858	\\
Abell 262	&	0.0162	&$	01	:	52	:	46.48	$&$	+36	:	09	:	06.53	$&	4823	\\
Abell 3581	&	0.0220	&$	14	:	07	:	29.50	$&$	-27	:	01	:	07.00	$&	6531	\\
Abell 4038	&	0.0291	&$	23	:	47	:	45.11	$&$	-28	:	08	:	26.67	$&	8605	\\
Abell 2634	&	0.0314	&$	23	:	40	:	00.84	$&$	+27	:	08	:	01.37	$&	9256	\\
Abell 496	&	0.0327	&$	04	:	33	:	37.84	$&$	-13	:	15	:	43.04	$&	9651	\\
Abell 2063	&	0.0341	&$	15	:	23	:	05.30	$&$	+08	:	36	:	33.18	$&	10055	\\
Abell 2052	&	0.0350	&$	15	:	16	:	44.50	$&$	+07	:	01	:	17.00	$&	10314	\\
Abell 2147	&	0.0354	&$	16	:	02	:	17.00	$&$	+15	:	58	:	28.25	$&	10419	\\
Abell 576	&	0.0381	&$	07	:	21	:	30.24	$&$	+55	:	45	:	41.69	$&	11200	\\
Abell 3571	&	0.0386	&$	13	:	47	:	28.39	$&$	-32	:	51	:	54.02	$&	11353	\\
Abell 2589	&	0.0412	&$	23	:	23	:	57.41	$&$	+16	:	46	:	37.94	$&	12088	\\
Abell 2657	&	0.0421	&$	23	:	44	:	57.42	$&$	+09	:	11	:	35.39	$&	12361	\\
Abell 119	&	0.0446	&$	00	:	56	:	16.10	$&$	-01	:	15	:	19.77	$&	13080	\\
Abell 3558	&	0.0470	&$	13	:	27	:	56.88	$&$	-31	:	29	:	43.71	$&	13762	\\
Abell 1644	&	0.0488	&$	12	:	57	:	11.58	$&$	-17	:	24	:	34.47	$&	14267	\\
Abell 3562	&	0.0502	&$	13	:	33	:	34.74	$&$	-31	:	40	:	20.16	$&	14677	\\
Abell 4059	&	0.0504	&$	23	:	57	:	00.40	$&$	-34	:	45	:	32.00	$&	14740	\\
Abell 3391	&	0.0552	&$	06	:	26	:	20.45	$&$	-53	:	41	:	35.89	$&	16102	\\
Abell 85	&	0.0554	&$	00	:	41	:	50.45	$&$	-09	:	18	:	11.46	$&	16138	\\
Abell 133	&	0.0560	&$	01	:	02	:	41.77	$&$	-21	:	52	:	55.75	$&	16314	\\
Abell 3158	&	0.0581	&$	03	:	42	:	52.95	$&$	-53	:	37	:	52.69	$&	16899	\\
Abell 3266	&	0.0610	&$	04	:	31	:	13.31	$&$	-61	:	27	:	11.43	$&	17724	\\
Abell 1795	&	0.0613	&$	13	:	47	:	22.56	$&$	+26	:	22	:	51.91	$&	17815	\\
Abell 2029	&	0.0779	&$	15	:	10	:	56.10	$&$	+05	:	44	:	41.19	$&	22447	\\
Abell 2142	&	0.0908	&$	15	:	58	:	20.03	$&$	+27	:	14	:	00.06	$&	25993	
\enddata
\tablenotetext{}{Notes.}
\tablenotetext{a}{the los velocity of BCG.}
\end{deluxetable}

%% file: Table2.tex
\begin{deluxetable*}{lrrcccccrc}
        \tablecolumns{10}
        \tabletypesize{\footnotesize }
        \tablewidth{2\columnwidth}
        \tablecaption{\label{tab:clusters}
        Properties of 29 HIFLUGCS clusters
        }
        \tablehead{
         \multicolumn{1}{c}{Name} &
         \multicolumn{1}{c}{D\tablenotemark{a}} &
         \multicolumn{1}{c}{$r_{500}$\tablenotemark{b}} &
         \multicolumn{1}{c}{$kT_{\rm vir}$\tablenotemark{c}} &
         \multicolumn{1}{c}{$\log(M_\mathrm{gas})$\tablenotemark{d}} &
         \multicolumn{1}{c}{$\log(M_\mathrm{bar})$\tablenotemark{e}} &
         \multicolumn{1}{c}{$\sigma_\mathrm{last}$\tablenotemark{f}} &
         \multicolumn{1}{c}{$r_\mathrm{last}$\tablenotemark{g}} &
         \multicolumn{1}{c}{$N_\mathrm{gal}$\tablenotemark{h}} &
         \multicolumn{1}{c}{Reference} \\
         \colhead{}
         &
         \multicolumn{1}{c}{(Mpc)} &
         \multicolumn{1}{c}{(kpc)} &
         \multicolumn{1}{c}{(kpc)} &
         \multicolumn{1}{c}{($M_{\odot}$)} &
         \multicolumn{1}{c}{($M_{\odot}$)} &
         \multicolumn{1}{c}{(km/s)} &
         \multicolumn{1}{c}{(kpc)}
         & &
         }
         \startdata
NGC 4636	&	13	&	255	&	0.9	&	11.230	$\pm$	0.059	&	11.379	$\pm$	0.056	&	229	$\pm$	21	&	29	$\pm$	7	&	234	& 28,29,30		\\
Fornax	&	20	&	250	&	1.3	&	11.724	$\pm$	0.006	&	11.827	$\pm$	0.018	&	252	$\pm$	36	&	1384	$\pm$	313	&	68	& 31,32		\\
Abell 3526	&	41	&	761	& 3.9	&	13.037	$\pm$	0.040	&	13.074	$\pm$	0.037	&	890	$\pm$	110	&	779	$\pm$	509	&	97	& 3,11,12,13,18,26,27		\\
Abell 1060	&	55	&	714	& 3.2   &	12.934	$\pm$	0.048	&	12.974	$\pm$	0.044	&	634	$\pm$	65	&	389	$\pm$	359	&	186	& 3,11,12,13,14,15		\\
Abell 262	&	68	&	755	& 2.4	&	13.033	$\pm$	0.048	&	13.070	$\pm$	0.044	&	551	$\pm$	68	&	2125	$\pm$	995	&	95	& 3,11,12,13,33,34		\\
Abell 3581	&	92	&	606	& 2.0	&	12.708	$\pm$	0.050	&	12.756	$\pm$	0.045	&	436	$\pm$	77	&	649	$\pm$	520	&	32	& 3,11,12,18,19,20,35,36	\\
Abell 4038	&	120	&	858	& 3.1	&	13.253	$\pm$	0.034	&	13.284	$\pm$	0.032	&	773	$\pm$	125	&	513	$\pm$	128	&	53	& 3,11,13,18		\\
Abell 2634	&	129	&	789	& 3.2	&	13.140	$\pm$	0.028	&	13.174	$\pm$	0.026	&	731	$\pm$	103	&	1220	$\pm$	340	&	73	& 3,11,13		\\
Abell 496	&	135	&	967	& 4.9	&	13.446	$\pm$	0.019	&	13.472	$\pm$	0.018	&	648	$\pm$	71	&	805	$\pm$	161	&	126	& 3,8,10		\\
Abell 2063	&	140	&	874	& 3.8	&	13.301	$\pm$	0.020	&	13.330	$\pm$	0.019	&	779	$\pm$	120	&	844	$\pm$	311	&	63	& 3,9,11,18,20	\\
Abell 2052	&	144	&	857	& 3.4	&	13.290	$\pm$	0.024	&	13.320	$\pm$	0.023	&	475	$\pm$	61	&	622	$\pm$	157	&	59	& 3,9,18,19		\\
Abell 2147	&	145	&	1057	& 4.1	&	13.592	$\pm$	0.044	&	13.615	$\pm$	0.042	&	811	$\pm$	106	&	1146	$\pm$	177	&	87	& 3,8,9		\\
Abell 576	&	156	&	869	&	4.1	&	13.301	$\pm$	0.154	&	13.330	$\pm$	0.144	&	923	$\pm$	126	&	1582	$\pm$	411	&	79	& 3,11,12,18,19,36		\\
Abell 3571	&	158	&	1133	& 7.0	&	13.713	$\pm$	0.024	&	13.734	$\pm$	0.023	&	841	$\pm$	84	&	999	$\pm$	536	&	99	& 3,11,18		\\
Abell 2589	&	168	&	837	&	3.9	&	13.248	$\pm$	0.029	&	13.279	$\pm$	0.027	&	610	$\pm$	83	&	1148	$\pm$	225	&	77	& 1,3,7,8,23 \\
Abell 2657	&	171	&	820	&	3.5	&	13.215	$\pm$	0.019	&	13.247	$\pm$	0.018	&	789	$\pm$	131	&	666	$\pm$	214	&	35	& 3,17,18	 	\\
Abell 119	&	181	&	1062	& 5.7	&	13.630	$\pm$	0.020	&	13.652	$\pm$	0.019	&	648	$\pm$	51	&	1240	$\pm$	316	&	240	& 3,6,7,8,9		\\
Abell 3558	&	190	&	1188	& 5.0	&	13.806	$\pm$	0.013	&	13.825	$\pm$	0.012	&	820	$\pm$	86	&	1558	$\pm$	781	&	138	& 1,3,11,12,27		\\
Abell 1644	&	197	&	1077	& 5.1	&	13.626	$\pm$	0.063	&	13.649	$\pm$	0.060	&	901	$\pm$	86	&	1060	$\pm$	120	&	211	& 1,3,7,19,37   	\\
Abell 3562	&	202	&	904	&	4.4	&	13.377	$\pm$	0.016	&	13.405	$\pm$	0.015	&	729	$\pm$	129	&	1269	$\pm$	267	&	48	& 3,8,27		\\
Abell 4059	&	203	&	892	&	4.2	&	13.350	$\pm$	0.033	&	13.378	$\pm$	0.031	&	666	$\pm$	94	&	926	$\pm$	301	&	50	& 1,3,8,19		\\
Abell 3391	&	221	&	971	&	5.8	&	13.500	$\pm$	0.026	&	13.525	$\pm$	0.025	&	885	$\pm$	152	&	1815	$\pm$	511	&	49	& 3,19		\\
Abell 85	&	222	&	1217	& 6.0	&	13.824	$\pm$	0.021	&	13.843	$\pm$	0.020	&	934	$\pm$	74	&	1536	$\pm$	300	&	465	& 1,2,3,4,5		\\
Abell 133	&	224	&	885	&	4.0	&	13.364	$\pm$	0.019	&	13.392	$\pm$	0.018	&	803	$\pm$	147	&	1021	$\pm$	191	&	45	& 1,3		\\
Abell 3158	&	232	&	1013	&	5.0	&	13.574	$\pm$	0.038	&	13.598	$\pm$	0.036	&	985	$\pm$	83	&	1581	$\pm$	340	&	206	& 1,3,7		\\
Abell 3266	&	243	&	1265	&	9.5	&	13.924	$\pm$	0.020	&	13.942	$\pm$	0.019	&	1226	$\pm$	95	&	2848	$\pm$	868	&	327	& 1,3,7,24,25		\\
Abell 1795	&	244	&	1085	&	6.1	&	13.695	$\pm$	0.012	&	13.716	$\pm$	0.011	&	831	$\pm$	99	&	3087	$\pm$	488	&	105	& 1,3,7,9		\\
Abell 2029	&	304	&	1247	&	8.3	&	13.916	$\pm$	0.015	&	13.934	$\pm$	0.014	&	844	$\pm$	41	&	2751	$\pm$	404	&	1056	& 16,17,38,39		\\
Abell 2142	&	349	&	1371	&	8.4	&	14.127	$\pm$	0.003	&	14.142	$\pm$	0.003	&	1062	$\pm$	70	&	3012	$\pm$	131	&	994	& 9,21,22		
\enddata
\tablenotetext{}{Notes.}
\tablenotetext{a}{the angular diameter distance by the cluster center as BCG;}
\tablenotetext{b}{the cluster radius in \cite{Zhang11};}
\tablenotetext{c}{the cluster viral temperature in \cite{Hudson10};}
\tablenotetext{d}{the X-ray gas mass of galaxy cluster within r$_{500}$ in \cite{Zhang11};}
\tablenotetext{e}{the baryonic mass of galaxy cluster within r$_{500}$ including the stellar mass estimated by the scaling relation in \cite{Giodini09};}
\tablenotetext{f}{the los velocity dispersion at the last binned point;}
\tablenotetext{g}{the projected radius of the last binned point;}
\tablenotetext{h}{the number of member galaxies.}
\tablerefs{(1) \citet{2016AJ....151...78P};
(2) \citet{2016MNRAS.458.1590A};
(3) \citet{2004AJ....128.1558S};
(4) \citet{2009ApJ...707.1691A};
(5) \citet{2015MNRAS.454.2050F};
(6) \citet{2012RAA....12.1381T};
(7) \citet{2009A&A...495..707C};
(8) \citet{2003AJ....125..478L};
(9) \citet{2011ApJ...736...21S};
(10) \citet{2008A&A...486...85C};
(11) \citet{2001MNRAS.327..265H};
(12) \citet{2001MNRAS.327..249S};
(13) \citet{1989ApJS...69..763F};
(14) \citet{2008A&A...486..697M};
(15) \citet{2011A&A...531A...4M};
(16) \citet{2013ApJ...773...86T};
(17) \citet{2008ApJ...687..899R};
(18) \citet{2000MNRAS.313..469S};
(19) \citet{2014ApJ...797...82L};
(20) \citet{2009AJ....137.4795C};
(21) \citet{2011ApJ...741..122O};
(22) \citet{2007MNRAS.379..867V};
(23) \citet{2011AJ....141...99L};
(24) \citet{2009ApJ...693.1840B};
(25) \citet{2017MNRAS.468.2645D};
(26) \citet{2006AJ....132..347C};
(27) \citet{2009Sci...326.1379C};
(28) \citet{2009MNRAS.400.1962K};
(29) \citet{2010ApJ...709..377P};
(30) \citet{2006A&A...459..391S};
(31) \citet{1989AJ.....98..367F};
(32) \citet{1990AJ....100....1F};
(33) \citet{2008A&A...486..755T};
(34) \citet{2012ApJS..199...36S};
(35) \citet{2009A&A...499..357G};
(36) \citet{2012ApJS..199...23H};
(37) \citet{2017A&A...599A..83M};
(38) \citet{2017ApJS..229...20S};
(39) \citet{2018AJ....156..224J}.
}
\end{deluxetable*} 